# Pitfalls and Protocols in Practice of Manufacturing Data Science

Chia-Yen Lee    Chen-Fu Chien

**Abstract**

The practical application of machine learning and data science (ML/DS) techniques present a range of procedural issues to be examined and resolve including those relating to the data issues, methodologies, assumptions, and applicable conditions. Each of these issues can present difficulties in practice; particularly, associated with the manufacturing characteristics and domain knowledge. The purpose of this paper is to highlight some of the pitfalls that have been identified in real manufacturing application under each of these headings and to suggest protocols to avoid the pitfalls and guide the practical applications of the ML/DS methodologies from predictive analytics to prescriptive analytics.

**Index Terms**: data science, manufacturing practice, machine learning, big data, prescriptive analytics

## 1. Introduction

According to the Fourth Industrial Revolution, often called Industry 4.0, machine learning technologies and data science analytics (ML/DS hereafter) will take on increasingly important roles as automation transforms on global supply chains and smart factory [1] [2]. In fact, manufacturing-process innovation is more critical [3] and ML/DS provides potential solutions to drive the technology migration. ML/DS techniques show the strengths, weaknesses and major functionalities to address difficulties and challenges in production systems [4]. Most of the Industry 4.0 literature has developed numerous ML/DS techniques for automation and prediction; however, few studies have investigated the practical aspects when applying ML/DS to manufacturing systems. This study fills the gap in literature and discusses several common issues (pitfalls hereafter) and the corresponding possible solutions (protocols hereafter), particularly, ad hoc methods based on manufacturing practice.

To make a success of the manufacturing data science, it is critical to understand the characteristics, data issues, and management challenges of the production systems, as shown in Table 1. All the data collected from the real-time sensors in the shop floor present the dynamics of the production characteristics. An excellent data scientist needs to devote himself/herself into the fab and understand the factory dynamics and human nature, and thus he/she can interpret the real issues from data log by investigating the bottom of the problem and provide insightful suggestion for improvement. We take two data issues and two management issues as examples, respectively.

For data issues, one example is about one categorical variable with too many categories (i.e. levels). Such variable like recipes, materials, or parts in high-tech industry showing categories more than one or ten thousands is common due to a variety of product types. The curse of dimensionality or computational burden arise as the analyst transforms them into dummy/binary variables. In another example, for R&D product or engineering-trial request, these types of products set some specific parameter or recipe for product development or small-volume multiple-type purpose. They may bring the data with outlier or noise to make the ML/DS model training unstable. They also commonly arise "one-shot trial" or the "small data" issue, i.e. the number of samples is insufficient, when compared with mass production bringing big data. Datasets that are too small can also result in skewed models or inadequate analytical outcomes. Big data is for population estimation while small data is for causal validation. In addition, these R&D-type products usually lead to frequent equipment setups and occupy some equipment used for general products of mass production. The capacity loss or equipment contamination could be potential problem.

Table 1 Production characteristics, data issues, and management challenges encountered by data analysts

| Characteristics | Data Issues and Management Challenges | Refer. |
|---|---|---|
| Batch production | Lot ID, merge or split, mixed lot, lot tracing | [5] |
| Small-volume multiple-type | Class/data imbalance problem, small samples for each specific product, changeover | [6] |
| Parallel machine | Missing value, identical or non-identical, old or new, high dimension, multicollinearity, scheduling complexity | [7] [8] |
| Bottleneck machine | Low throughput, Little's law, variability improvement for line balance | [9] |
| Machine capability | New/old machines with different throughputs, process supports, utilization, work-in-process (WIP), part replacement frequency, etc., class/data imbalance problem | [10] |
| Recipe and parts | Too many levels in one categorical variable, dummy variables transformation from categorical variable, high dimension | [8] |
| Sampling testing | Missing value, multi-response, metrology delay | [11] |
| Engineering or R&D lot | Design of experiments with small datasets, outlier, setup time, machine occupied (capacity loss), machine failure | [12] |
| Maintenance | Capacity loss, scheduled or non-scheduled downtime, mean-time-to-repair (MTTR), mean-time-between-failure (MTBF), manual check list, text, typing error, majority class "Others" regarding failure/root cause | [13] |
| Inventory | Different types of inventory, Inventory = Lead Time + Uncertainty | [14] [15] |
| Changeover | Capacity loss, sequence-dependent setup time | [16] |
| Bottleneck shift | Change of product-mix, different treatment, WIP transfer | [17] |
| Queue time limit | Process route, transportation, batch production, defects, WIP | [18] [19] |

For management challenges, the bottleneck machine which is one process with limited capacity and low throughput is critical in the manufacturing shop floor. The bottleneck affects supply overstock, production cycle time, and work-in-process (WIP) directly. The Little's law shows a typical relationship among cycle time, WIP, and throughput for factory management based on queueing theory [9]; that is, the cycle time deteriorates exponentially when WIP increases over a capacity limit [20]. Bottleneck also affects the line balancing from upstream and downstream aspects, and thus we not only improve the throughput of bottleneck but also need to improve the variability (i.e. reliability) of the bottleneck such as improvements of mean-time-between-failure (MTBF) and mean-time-to-repair (MTTR). Take inventory issue as another example, inventory including materials, parts, WIP, finished goods, etc. is one kind of "muda" in a manufacturing system [21]. Excess inventory will result in obsolescence or loss from falling price. Generally, inventory is a "result" rather than a "cause". The literal equation " $Inventory = LeadTime + Uncertainty$ " is useful and helpful to keep in mind for inventory reduction. It implies that shortening the production lead time and reducing the uncertainty are the insightful ways to reduce inventory essentially [22] [14]. For example, scheduling is a typical method for cycle time reduction, and predictive analytics helps uncertainty elimination by data collection and ML/DS techniques.

In practice, ML/DS provides a variety of applications for analyzing manufacturing systems. Generally, data analysts prefer to use of prediction models with higher accuracy (i.e. minimal mean squared error (MSE)) via cross validation [23]. For accuracy, the neural network models [24] [25], kernel methods, [26] or ensemble methods such as random forest [27] and boosting [28] have been developed, and while these techniques can improve accuracy dramatically and address overfitting issues well, they are difficult to be interpreted [29]. In many real applications, *interpretation* is more attractive than prediction accuracy for clarifying the scientific causal relationship rather than a statistical correlation. For example, when management needs to consider a variety of decision risks, instead of using MSE, showing the mean absolute percentage error (MAPE) of a prediction model is more meaningful. For example, choosing the best answer may not be intuitive in the case of "Prediction Model A shows 95% accuracy with a big loss if misclassified, and Prediction Model B shows a 90% accuracy with a small loss if misclassified." If management considers the decision risks, the story changes and pushes the predictive thinking toward the prescriptive analytics, i.e., the focus is now on the *tradeoff of decision risks* and *resource optimization*.

This study is motivated by the production characteristics, data issues, management challenges, and decision risks from a practical aspect. To promote ML/DS and automation, we list several pitfalls and their corresponding protocols in practical manufacturing systems.

## 2. Pitfalls and Protocols

This section describes the solutions (i.e. protocols) to 12 common pitfalls encountered when applying ML/DS to manufacturing systems.

**Pitfall 1**. Can ML/DS identify the important variables/features? The feature selection technique identifies or extracts the minimally sized subset of features from a database that (1) improve prediction performance (accuracy); (2) provide simpler, faster, and more cost-effective predictors (fewer control charts for process monitoring); (3) approach the original class distribution, given only the selected variables; (4) and provide a better understanding of the causal relation/physical meanings among variables [30] [31]. In literature, previous studies typically mention two types of feature selection techniques: variable selection and feature extraction (the latter is also called variable transformation). Variable selection selects the best subset of the raw variables without a transformation and generally provides the supervised learning with labels, such as stepwise selection [32] [33], least absolute shrinkage and selection operator (LASSO) [34], classification and regression trees (CART) [35], random forest [27], and Boosting [36] [37]. Feature extraction transforms the raw variables into a lower dimensional space and generally is the unsupervised learning without labels, such as principal component analysis (PCA) [38], independent component analysis (ICA) [39] [40], Ward's clustering [41], and K-means clustering [42]. For simplicity, the variable is selected from raw input columns; however, the feature is constructed from the combination of raw input columns [30]. This paper uses "variable" rather than "feature" when there is no impact/confusion on the selection algorithms. In manufacturing practice, variable selection is more commonly used than feature extraction since the feature (e.g., the principal component) constructed by several raw input variables is not easily understood or is difficult to interpret, even though the extracted features can improve the prediction accuracy. However, the use of variable selection can introduce other issues, such as whether only one ML/DS variable selection technique can identify the important variables or whether ML/DS can effectively identify the important variables.

**Protocol 1**. Using only one ML/DS to identify the important variables is risky, particularly when we only use the linear stepwise selection model to investigate the main effect of each individual predictor without considering the higher interaction effect among the variables. In fact, it is difficult to understand the geometric relation (linear or nonlinear) among variables from the collected dataset because the manufacturing dataset involves a complicated process network with the interaction effect among several processes, and thus we have no idea how to pick the right feature selection technique. Therefore, to select the most robust variables, we can use multiple techniques with linear and nonlinear models simultaneously. Voting is a common method because it ranks the variables by the sum of their selected times, using different feature selection techniques [8]. Table 1 gives an illustration, where 1 indicates if the variable is selected; otherwise 0.

Table 1 Variable selected by voting

| | Stepwise selection | Lasso | Random forest | Boosting | # of votes received |
|---|---|---|---|---|---|
| Var_108 | 1 | 1 | 1 | 1 | 4 |
| Var_32 | 1 | 1 | 1 | 0 | 3 |
| Var_79 | 0 | 1 | 1 | 1 | 3 |
| Var_50 | 1 | 0 | 1 | 1 | 3 |
| Var_53 | 1 | 0 | 0 | 1 | 2 |
| Var_14 | 1 | 1 | 0 | 0 | 2 |
| … | … | … | … | … | … |

Can ML/DS really work to identify the important variables? The answer appears to be "it depends". For example, after collecting data including sensors and parameters from manufacturing equipment, if we see one column with identical values (i.e., all observations have the same value in this variable), based on the ML/DS there is no information provided to distinguish the observations, and thus the variable is unimportant or statistically insignificant. Even so, this variable can be very important (e.g., equipment developer instructed this parameter is so critical and no adjustments is allowed after equipment installation). In this case, ML/DS may not successfully identify this parameter due to the column with identical values.

**Pitfall 2**. Can put all raw variables into feature selection technique?

If a dataset contains many raw variables (e.g., less than 300), we may put them into a feature selection technique and generally obtain the result in a reasonable run time. However, the number of raw variables can exceed 1,000 or 10,000, such as in predictive manufacturing [1], semiconductor manufacturing [43], gene expression [44], or bioinformatics [45], and the run time can be excessive, and even run out of memory.

**Protocol 2**. For a large amount of variables, we can remove the unimportant variables and then apply the feature selection technique. Based on the sufficient and necessary condition, if there is a causal relationship between two variables, then they must show some degree regarding the correlation between them. Thus, we can derive "if there is a no correlation between two variables, they must show no causal relationship" and build a quick filter. We can pairwise calculate the Pearson's correlation coefficient, the Mann-Whitney U Test, the Chi-Square Test of Independence for categorical variables, between a large number of predictors and only one response variable for preliminary filtering, and remove the predictors if the absolute value of the correlation coefficient is smaller than some threshold. These simple statistics can be calculated very quickly and filter out a bunch of uncorrelated variables quickly. However, using correlation test to remove uncorrelated predictors assumes that we consider only the main effect of each individual predictor and ignore the interaction effect among the predictors. In practice, we can also conduct the design of experiment (DOE) [46] to confirm the main effect of individual variable and the interaction effects among multiple variables after the quick filter.

**Pitfall 3**. Does the selected variable not show the physical causal relation?

That is, data-driven ML/DS approach investigates the "correlation" among variables through the statistic calculation (eg. correlation coefficient) or model training/fitting process (eg. ordinary least square, OLS) to identify the significant variable. The correlation built by ML/DS does not imply the causal relation between variables particularly in the physical or chemical sciences. Thus, there statistically-selected variables may not be interpretable via engineering validation.

**Protocol 3**. The feature selection should be an iterative procedure between data scientists and engineering validation. Each iteration provides the selected variables for engineering validation, and we can remove some of the variables without physical meanings, and then re-run the feature selection technique with the remaining selected variables, re-identify the important variables and re-send them for engineering validation. In general, three iterations are sufficient for the selected variables to converge. Note that if there are so many variables selected (more than 100) in the first round, data scientists may provide the number of selected variables less than 20-30 to engineers in each round since it may take time to investigate the causal relationship by doing experiments or calling suppliers. When we cannot confirm a selected variable by physical checking, we can keep it for the next iteration. Finally, the knowledge management (KM) of these selected features is suggested to for future tutoring.

**Pitfall 4**. How to enhance the interpretability between predictors and response variable?

For a complicated nonlinear data pattern, we can use the support vector machine (SVM) [47] /neural network [48] [24] /deep learning [25] to build the function/relationship between predictors and response variable. While these powerful ML/DS techniques can improve the prediction accuracy, they can also undermine the interpretability of the relationship between predictors and response variable (i.e. a black box) [49].

**Protocol 4**. To address the issue, we may ask a question first: is it necessary to make all ML/DS techniques explainable in real applications? In some cases, we may just treat ML/DS as a module/unit/subfunction or a small part in a whole analysis flow, and thus emphasize input and output of the module rather than how does it process inside the module. Recently, the explainable artificial intelligence (XAI) arises and is helpful to enhance the interpretability [50] [51]. Another useful method, so-called "divide-and-conquer" strategy [52] [53], can be suggested to dissolve a complicated nonlinear data pattern. In the "divide" phase, we decompose the data pattern into several relatively simple or regular sub-patterns, and in the "conquer" phase, we can build several simple models/weak classifiers to fit each sub-pattern and improve the interpretability, respectively. The following examples of time-series panel datasets suffice. For the signal processing, we can apply the empirical mode decomposition (EMD) and Hilbert spectral analysis to generate a collection of intrinsic mode functions (IMFs) [54]. These IMFs are practically orthogonal and present the instantaneous frequencies regarding the local properties of the data we can then use to explore the physical interpretations of the nonlinear non-stationary dataset. For economics or statistics, we can decompose a time-series panel data (e.g., oil price over past several decades) into four sub-patterns: "Trend", "Cyclic (long cycle)", "Seasonal (short cycle)" and "Random Noise" [55]. Figure 1 shows a "detrending" time series (weekly) dataset of Brent Oil Prices from June 26, 1988 to March 31, 2019. Then we can suggest some simple models to fit these sub-patterns, such as characterizing "Trend" by regression or SVM, "Cyclic/Seasonal" by autoregressive integrated moving average (ARIMA) [56], generalized autoregressive conditional heteroskedasticity (GARCH) [53], or some trigonometric functions, and "Noise/Residual" by neural networks. The feature selection can also be applied to these sub-patterns. Other techniques for improving the interpretability include clustering [57], time series segmentation [58] [59], and sliding window/moving average filter [60].

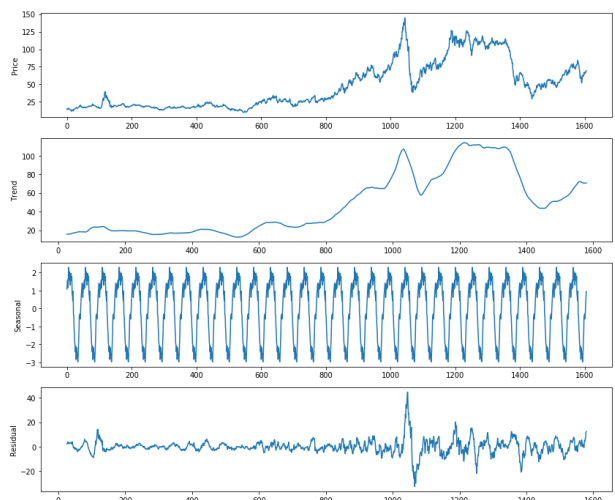

Figure 1 Time series decomposition of oil prices 1988-2019

**Pitfall 5**. How to handle when transforming categorical variable into too many dummy variables?

Dummy variables are binary variables which can be generated from the categorical variable and represent mutually exclusive categories. Each of the possible values of a categorical variable is referred to as a level. Typically, we can transform one categorical variable with $L$ levels into $L-1$ dummy variables to avoid the multicollinearity issue [61]. In a categorical variable, its one dummy variable represents some categorical-level effect of absence or presence which is expected to shift the outcome upward or downward. Not every dataset with categorical variables needs dummy variable transformation before we build it into the classification/prediction models; "it depends" on the underlying assumptions or characteristics of the classification/prediction models we use. Some ML/DS packages can handle the categorical variables well and automatically in the classification/prediction models. However, if one categorical variable has a very large number of levels $L$ (e.g., a recipe or parts in some manufacturing systems), then too many dummy variables could introduce the curse of dimensionality, i.e., "for model training process the number of observations required exponentially grows to estimate the function or model parameters" [26].

**Protocol 5**. To avoid generating too many dummy variables, three tips are suggested in practice. First, we can use concept hierarchy, which can reduce the data by grouping and replacing low-level concepts with high-level concepts, or combining several levels with a similar concept into one higher level [62], such as merging tools into tool groups or merging products into product groups. Second, we can shorten the data-collection periods, which can reduce the number of dummy variables generated from the categorical variable (eg. many types of recipes in one year) challenge the model training process; in particular, when the mix of products change several times to meet customer demand fluctuations. This unstable nature will undermine the prediction accuracy. Third, we can remove the level shown only once in the categorical variable, based on the idea that the level cannot be repeatable or re-validated. In practice, there are many levels that are shown only once which can confuse the model training and reduce the prediction accuracy (e.g., special recipes generated by an engineering experiment or pilot-runs of a special product). From a scientific aspect, these levels do not provide any evidence of repeatability and reproducibility, and thus they cannot be used to train the ML/DS model.

**Pitfall 6**. Many missing values in one variable/column (or observation/row)

The missing value imputation technique is based on finding "the relationship from other columns (or rows) for imputing missing value". For details and methods, see [63] and [64]. Here, we focus on if there are many missing values in one column (or raw) (eg. missing over 50%).

**Protocol 6**. Generally, if engineering validation confirms that a variable is important, it is rare to see many missing values in this column. Once we see some variables with many missing values (e.g., over 50%), we can remove the column since filling out 50% missing values is unreliable. However, there is an example, if one variable $X_{70}$ with many missing values is removed but it is highly correlated with $X_{14}$ which has no missing values (i.e., using existing values in $X_{70}$ calculates the correlation coefficient between these two variables). After we use feature selection, if $X_{14}$ is selected as an important variable but with poor interpretation, we need to check the causal relation/physical meaning of $X_{70}$.

**Pitfall 7**. Merging data tables and handling many missing values after the merge.

Since information systems and sensors collect raw data in diverse formats (e.g., analog and digital signals, sampling rates, etc.) across many divisions, data analysts have developed different methods to manage the continuous streams of big data. In the data merge process, we can identify the "key" (such as LotID, MachineID) or "composite key" (such as LotID_Date_Time_Recipe combining four variables) and use them to integrate the data of interest into one table. Generally, the data merge integrates two types of data tables: event-based records and monitoring-based records. The event-based dataset records when an event triggers, and the monitoring-based dataset records periodically. In the data preprocessing, we can merge these two types of data tables into a one table for easier analysis. However, the data recording mechanisms are very different in these two, which one should be the "main" table to concatenate the other type of table?

**Protocol 7**. Which table should be the main table? It depends on the problem we need to solve. Generally, we use the event-based record as the main table when we are troubleshooting, because we want to find the event that causes a machine failure. We use the monitoring-based record as the main table when we need to analyze a machine's operation over a defined period for facility/energy/process monitoring. Table 2 lists the differences between the two records.

If the event-based record is our main table, we can concatenate the data by nearest time, roll-forward, or roll-backward. Nearest time indicates that the new variable in the other table is merged into the main table by the corresponding key with respect to the closest time, regardless of what occurs before/after the event. Roll-forward merges the new variable in the other table into the main table before the event occurs (i.e., it rolls the time of the new variable forward to meet the event-based record, and we fill out the closest record in the past).; Roll-back merges the new variable in the other table into the main table immediately after the event.

Table 2 Comparison of event-based and monitoring-based data

|  | Event-based Record as Main Table | Monitoring-based Record as Main Table |
|---|---|---|
| Data recording | Event triggers recording the data | Record data periodically |
| Example | Equipment parameter adjustment/tuning | Temperature sensing per second |
| Table characteristics before merge | Records with relatively fewer or sparse samples | Records with a more complete dataset |
| Pros after merge | Fewer missing values | Observe the periodic change |
| Cons after merge | No data in a long period | Many missing values |
| Methods | Nearest time Roll-forward/Roll-back | Nearest time Roll-forward/Roll-back |
| Purpose | Troubleshooting | Process/facility/energy monitoring |

In practice, data table merges can also generate missing values. After removing the columns or rows by data preprocessing, the sample size could become fewer and may affect the model training process and its prediction performance. There is a tip suggested here. After using the processed data for feature selection and the important variables are selected, to increase the number of samples we can turn back and repeat the data merging process again with these important variables only. The results show that our merged table has fewer missing values because the important variables in each information system usually have relatively complete data in the past long-run development.

**Pitfall 8.** Does the multicollinearity problem matter?
The multicollinearity problem increases the variance of the coefficient estimate of one predictor in a multiple regression model because it can be linearly predicted from the other predictors. In this case, the coefficient is unstable and may change erratically in response to a small change in the dataset. It implies a difficulty in the interpretation of the coefficient. In a worst case, the multicollinearity problem causes a switch in the sign of a coefficient, which in turn leads to a model misspecification or model invalidation. We can detect the presence of multicollinearity by (1) observing a large variation of estimated coefficient when one predictor is added or removed; (2) calculating the variance inflation factor (VIF) which is larger than 10 for each individual variable or when the average VIF is larger than 6 for all predictors in a regression model [65] [66].

**Protocol 8.** Addressing multicollinearity problem depends on (1) your purpose of prediction and (2) the prediction model you choose. Basically, the fact that we ignore checking multicollinearity in ML/DS techniques isn't a consequence of the algorithm but it's a consequence of the goal. Since multicollinearity issue does affect the predictive power but bias the estimated coefficient in the regression, for the prediction purpose, we may not be interested in the coefficients but could put more focus on the loss function (e.g., mean squared error, MSE), AUC (i.e., area under the receiver operating characteristic (ROC) curve), or F1 score (i.e. the harmonic mean of precision and recall) [67], which significantly affect the prediction performance. If the interpretation of predictor is important to clarify the causal relation, the regression model is suggested. In such cases, we can apply the shrinkage method, also known as regularization technique (e.g., ridge regression [68] or LASSO [34] [50] which can reduce the variance of coefficients although at a slight increase in bias due to a bias-variance tradeoff. However, LASSO can compulsorily shrink the coefficients of the predictor to 0 by adjusting the penalty parameter and thus improve generalization.

In addition, different ML/DS models we choose can cause different degrees of influence by multicollinearity on the analysis results. For example, a random forest approach, which randomly resample via bootstrap and randomly select the variables, is applied to build several subsets of data and bag the decision tree by out-of-bag validation [27]. That is, randomization builds many trees to form a forest, where the trees can be constructed by classification and regression trees (CART) [35], Chi-square automatic interaction detector (CHAID) [69] [70] [57], or C4.5/C5.0 with information gain [71] [72], etc. The performance of branching considers the information theory or the purity in the child node, and thus "IF-THEN" rules are generated for interpretation without using the coefficients of predictors in regression. In this case, highly-correlated or collinear variables does not significantly undermine the performance of the random forest; in particular, we can directly remove one if two collinear variables provide the same purity of child nodes. Similar conclusion "multicollinearity does not affect the model performance significantly" can be applied to such as SVM, neural networks, or deep learning, which put more focus on the prediction accuracy rather than the interpretability.

In practice, removing highly-correlated variables can effectively reduce the number of predictors/neurons used in the model (eg. neural network/deep learning) and then consequentially reduce the number of parameters which need to be estimated to avoid the curse of dimensionality. In most practical cases, multicollinearity is not always a problem if the prediction model shows excellent accuracy and robustness. But it deserves an investigation when we attempt to identify a correct model, enhance the physical interpretation of the causal relation, and reduce the number of variables. After all, in practice, less variables we use, lower cost we spend for model maintenance and management.

**Pitfall 9.** Does a higher prediction accuracy support a better decision-making?
First, if we consider predicting a continuous value of a response variable, there is nothing related to right or wrong since it emphasizes the accuracy according to the R-squared or MSE. However, to classify an observation into distinct classes (i.e. labels) involves misclassification and generates a confusion matrix with a Type I error, also called false positive/false alarm, and a Type II error, also called false negative/miss. There is an example in the manufacturing process. If we collect the equipment parameters and sensor data to predict the continuous response variables about the quality, yield, precision, thickness, metrology measure, etc., then in this case, the prediction result is only related to accuracy. However, once we can add the upper control limit (UCL) and lower control limit (LCL) for process monitoring or quality inspection, in this case Type I (claim out-of-control but truly in-control) and Type II (claim in-control but truly out-of-control) may occur. Figure 2 illustrates the example.

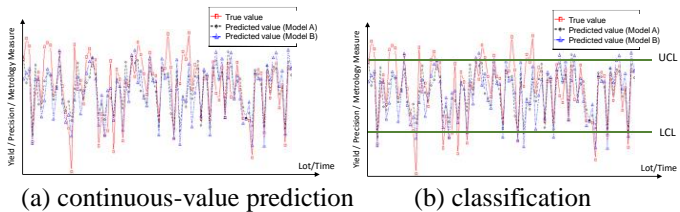

(a) continuous-value prediction　(b) classification
Figure 2 Distinguish prediction from classification

Either error type can affect the decision-making process by assessing the "decision risk". Figure 3 shows a manufacturing example of building two prediction models for classification of 128 lots. The performance metrics include accuracy, recall/sensitivity, specificity, prevision, F1-score, and AUC (area under ROC curve). Based on these metrics, the result shows that Model B presents better accuracy than Model A, and thus Model B is used to predict the metrology measure and quality inspection. However, if we further investigate the two error types in the confusion matrices of the two models, we can find that there are fewer Type II errors in Model A is less than in Model B. In most manufacturing cases, the monetary value (i.e., cost) of a decision risk regarding a Type II error (i.e., prediction showing in-control but truly out-of-control) is higher than the risk of a Type I error because a Type II error will lead the manufacturer to ship the product and receive a customer complaint later. Therefore, the decision maker may prefer to choose Model A, which lose some accuracy, to offset a larger decision risk. In conclusion, a Type I error or a Type II error involves a tradeoff.

| Model A | | True | |
|---|---|---|---|
| | | Fail | Pass |
| Predicted | Fail | 61 | 29 |
| | Pass | 7 | 31 |

| Model B | | True | |
|---|---|---|---|
| | | Fail | Pass |
| Predicted | Fail | 47 | 7 |
| | Pass | 21 | 53 |

| Model | Prediction Performance | | | | | |
|---|---|---|---|---|---|---|
| | Accuracy | Recall/ Sensitivity | Specificity | Precision | F1 Score | AUC |
| Model A | 71.9% | 89.7% | 51.7% | 67.8% | 77.2% | 70.2% |
| Model B | 78.1% | 69.1% | 88.3% | 87.0% | 77.0% | 78.9% |

Figure 3 Classification results of two MS/DS models

**Protocol 9**. The prediction model (i.e., predictive analytics) and decision risk (i.e., prescriptive analytics) complement each other. As an example, we can build the ML/DS model to estimate the probability distribution of in-specification and out-of-specification with respect to product quality inspection, and find an in-specification with the probability 0.2 and an out-of-control with the probability 0.8. Now we need to decide whether to ship or to scrub/rework. Table 3 lists the decision risks. If the product is in specification, we can ship the product at a cost equal to $0, or scrub/rework at a rework cost including material or labor equal to $20. If the product is out of specification, we can ship the product and receive customer complaint at a cost equal to $200, or scrub/rework in house at a cost equal to $20. Based on the calculation of the expected costs, we finally decide to scrub/rework.

Table 3 Payoff/cost matrix of decision risk

| True condition | Ship product | Scrub/Rework |
|---|---|---|
| In-specification (with probability 0.2) | $0 | $20 (material/labor-hour for rework) |
| Out-of-specification (with probability 0.8) | $200 (customer complaint) | $20 |
| Expected cost | $160 | $20 |
| Decision: Scrub/Rework | | |

In fact, the Scrub/Rework decision does not totally depend on the probability of out-of-control 0.8, but also depends on the results of decision risk (i.e., payoff/cost). Take a counterexample, where the material usage and labor hour for rework costs $2000 rather than $20; should we still decide to Scrub/Rework? Based on this case, both the prediction of in-specification/out-of-specification (i.e., estimating the probability distribution) and the decision risk measure (i.e., payoff/cost) are important. The former calculates probability distribution by ML/DS and the latter investigates the payoff/cost by expert/domain knowledge/decision maker's preference/sensitivity analysis [73]. Thus, we can conclude that prediction (i.e. predictive analytics) and decision risk (i.e. prescriptive analytics) are complementary.

As we have all known that prediction cannot be totally correct, and thus we need to consider the decision risk. Predictive analytics uses the data to predict what will happen and prescriptive analytics uses it to prescribe what should be done [74]; in particular, operations research (OR) and optimization techniques are commonly used for prescriptive analytics. There is an example of prescriptive analytics. Lee and Chiang (2016), who empirically studied aggregate capacity planning in the TFT-LCD industry, proposed a two-phase framework consisting of demand forecasting in the phase 1 and capacity decision in phase 2. Figure 4 shows three demand forecast models for phase 1. Three models suggested increasing or flat demand forecasts of the technology node A in the future; however, demand drops truly. In Figure 4, if we see a short increasing trend just before the validation dataset, there is no one believing the dropping demand in the future; that is, the three prediction models are justified. To address the inaccurate demand forecasts, Lee and Chiang investigated the decision risks (or called the regrets in their study) and proposed three capacity decision models including expected value model, minimax regret model [75], and stochastic programming model [76] to develop a robust capacity plan for phase 2. Their results showed that the capacity decision corrected the poor demand forecast in phase 1, which and balanced between capacity-shortage risk and capacity-surplus risk; in particular, the suggested capacity plan was robust with little fluctuation.

In practice, prescriptive analytics provides three benefits to optimize decision making: (1) assessing decision risk and balancing the tradeoff between different types of risks (e.g.,

          

Type I error and Type II error); (2) solving other problems beyond the scope of predictive analytics such as multi-objective production scheduling [78], inventory management [15], material investigation [79], project management [80], vehicle routing [81], financial portfolio optimization [82], vendor selection and order allocation [83], and decision analysis [73]; and (3) considering limited resources (eg., men, machines, materials, methods, measurements, environment). Thus, prescriptive analytics complements predictive analytics to enhance the connection with decision-making process.

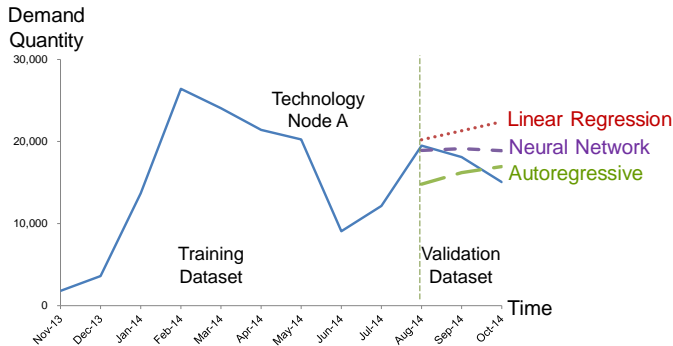

Figure 4 Three demand forecasting models [77]

**Pitfall 10**. How reliable is the conclusion derived from ML/DS? Is the conclusion or prediction results derived from ML/DS techniques reliable or usually true? How to criticize the conclusion derived from ML/DS? To answer the question, we need to know if ML/DS is an induction method or a deduction method. Induction is an inference process from specification (i.e., sample) to generalization (i.e., model) and deduction is an inference process from generalization to specification. In most ML/DS techniques, we can train a generalized model after investigating the patterns from each specific sample (i.e., induction), and then plug in the new data for prediction by using the generalized model, assuming the new data satisfy all of the model's underlying assumptions (i.e., deduction). Using a linear regression as an example, based on the dataset, the ordinary least squares (OLS) estimator in step 1 is an optimization applied to build a generalized linear equation which describes the pattern of the training dataset (i.e., nonparametric induction), and the regression line in step 2 predicts the new data (i.e., parametric deduction) according to all the underlying assumptions satisfied such as linearity relationship between predictors and response variable, independence of the errors, normality of the error distribution, homoscedasticity, and lack of perfect multicollinearity [84]. In fact, this two-step framework (i.e. induction for training and deduction for prediction) can generally applied to most of ML/DS techniques.

**Protocol 10**. We know that the conclusion derived from induction may not be true, whereas the result derived from deduction must be true if the given premise/assumption is true [85]. Here, we learn how to argue induction and deduction, respectively. For the induction step, since the induction is a transformation from specification to generalization, to criticize the result derived from induction we can argue the collected data/case/sample/observation, for example:

● Find the counterexample;

● Argue the representativeness of collected samples; and

● Argue the small/bias sample size.

Therefore, we focus on arguing the "justification of model training process"; in particular, the samples used for training. For example, a prediction model trained by the R&D products is problematic to be used to predict the case of normal products for mass production. For another example, given a different product-mix of every six months, it is doubtful that we can train a prediction model with the first six-months of dataset and then use it to predict the next six months.

For the deduction step, since the deduction is a transformation from generalization to specification, to criticize the result derived from deduction we can argue the assumption/premise, for example:

● Violate the assumptions in premise (which may be derived from induction, e.g., OLS);

● Find other causes which may derive the same conclusion;

● Argue the causal relationship (necessary conditions such as high correlation, sequence, and coherence)

Therefore, we focus on arguing the "justification of premise and causal relation"; in particular, the assumptions violated. For example, we cannot use linear regression for prediction if it violates linearity, independence, normality, homoscedasticity, etc. For example, high correlation does not imply causal relationship and usually ML/DS can only build a correlation without physical interpretation.

**Pitfall 11**. How to start the ML/DS works? How to collect dataset? How much data we need?
Clearly, the data we collect depends on the problem to be solved. In general, manufacturers do not maintain or collect specific datasets unless a problem arises.

**Protocol 11**. Rather than collecting massive amounts of data, we suggest collecting important dataset first, that is, an "event" which caused a big loss before. We can collect information about the problem in order to avoid the event occurring again. To collect data systematically, we can use an entity-relationship (ER) model to describe the interrelationship among the factors and use it to build the foundation of a specific domain knowledge [86]. It is usually used for the design of database management system (DBMS), i.e. relational database. In fact, it is important to update the E-R model when adding or deleting one column in the table of DBMS. In the long run lots of columns added or deleted without updating the E-R model can lead to a catastrophe for system integration in practice. Note that when a relational database suffers in a big data environment, we can suggest NoSQL (not only structured query language), also known an unstructured database, for parallel-distributed computing, schema free, or horizontally scalable system [87].

How much data we need? In fact, data heterogeneity is much more critical than big. Data heterogeneity not only helps us gain a better understanding of big data, but also provides a more comprehensive view leveraging data from a variety of real applications to enhance the prediction accuracy. In addition, data volume usually depends on the length of time interval in data collection process. Manufacturing systems tend to use long data-collection periods for investigating the "process nature" and building a "generalized" prediction model whose

  

parameters or coefficients of predictors can present a trend or a general response; however, a short-period data is used to build a "specific" prediction model which can present a more precise causal relation to support troubleshooting/process diagnosis in this specific period. This is a trade-off about "generalization versus specification" of building your prediction models.

How to start the ML/DS works for manufacturing system in the very beginning? Figure 5 shows three phases for improving manufacturing systems: Gemba Kaizen, Data Science, and System Integration, and each phase is characterized by Problem, Method, and Performance, respectively.

Gemba Kaizen emphasizes *observing, understanding, and improving* the manufacturing process in the shop floor. Gemba Kaizen investigates problems such as muda (i.e. waste), work-in-process (WIP), bottleneck, missing operation (MO), or labor capability. Lean production [21], work study [88], facility layout [89], and process improvement [90] are the principles, philosophies, and methods introduced for training labor and fully identifying the problem. In this phase we can develop the standard operating procedure (SOP), eliminate the waste (e.g., WIP or unnecessary motions), start collecting data and understand the shop-floor dynamics., formulate the performance metrics for management. In fact, this phase contributes to the benefits of understanding the characteristics of the manufacturing system and collecting dataset (even manually) prepared for next phase.

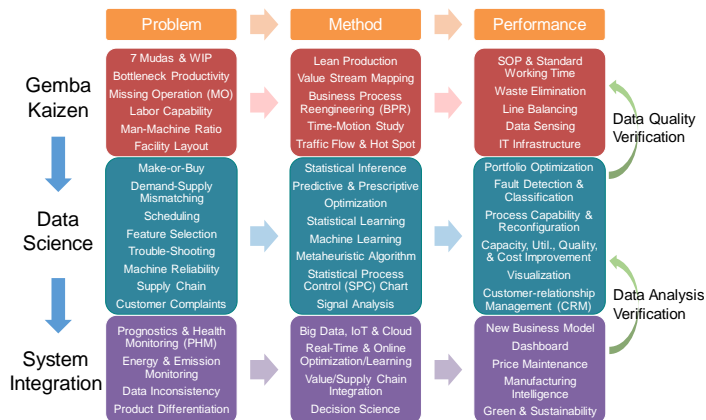

Figure 5 Three phases for improving manufacturing systems

The second phase Data Science emphasizes *predictive/prescriptive analysis* by using IT (information technology) and ML/DS techniques, and support the business process automation, operations automation, and engineering automation. Since parts of data system and IT infrastructure have been built in the previous phase, we collect and use data for analytics such as demand-supply mismatch, capacity planning, troubleshooting, scheduling, equipment reliability, customer complaints, etc.; in particular, to develop individual information system addressing each specific problem. Methods involve optimization [74], statistical process control (SPC) [46], machine learning [26], statistical learning and data mining [50], etc. In this phase we can build several information systems such as fault detection & classification (FDC), resource portfolio optimization, customer-relationship management (CRM), etc. to provide the analysis results and automation prepared for next phase usage. Due to garbage-in garbage-out, the performance generated from this phase significantly depends on the *data quality* from previous phase, and thus it is necessary to keep monitoring and verifying the data quality in Gemba Kaizen phase according to the results of ML/DS analysis.

The third phase "System Integration" emphasizes *synergy of manufacturing intelligence* by integrating various information systems, and supporting a comprehensive understanding of the manufacturing system for business sustainability. Since different types of analysis results are obtained from previous phase, the analysis results can be merged to support several comprehensive integrations such as facility energy consumption, emission & recycle monitoring, data inconsistency, prognostics & health monitoring (PHM) [13], product differentiation, etc. An integration needs the data and analysis results from multiple sources to present the "synergistic effect" through big data [1] [91], cyber-physical systems (CPS) [2], internet of things (IoT) [92] [93], cloud manufacturing [94], on-line learning and optimization [95], value and supply chain integration [96] [97] [98] [99], decision science [73], etc. In this phase we are inspired by the information technology (IT) and motivate manufacturing revolution such as new business model, dashboard of facility monitoring, price maintenance and cost advantage, green & sustainability, etc. to support the business strategy development and technology migration. Note that due to garbage-in garbage-out, the performance generated from this phase significantly depends on the data "analysis" from previous phase, and thus it is necessary to re-check/adjust/update the data science models and analysis results with newly dataset in previous "Data Science" phase according to the results of system integration.

**Pitfall 12.** How to develop roadmap and future works for smart factory?
What is the roadmap and future works for smart factory? What's the methodology we should use to improve the manufacturing systems? How to investigate the quality, yield, cost and profit when ramping the fab?

**Protocol 12.** The future of manufacturing continues to be studied intensively; see [1], [100], [2], [101] [102] for details. In practice, moving to a smart factory environment depends on the characteristics of each manufacturing system (e.g., IT infrastructure, educational level of labor, degree of automation, and product development). Besides to the roadmap like Figure 5, Figure 6 shows a three-phase product life cycle, neglecting the Decline phase, and the methodologies corresponding to yield and cost. Phase 1 (R&D) involves product development (eg. new-tape-out in high-tech industry), and the engineering parameter optimization and design of experiments (DOE) with limited datasets. Phase 2 (Growth) attempts to reduce the lead time for ramping-up the fab for product launch and time-to-market, particularly production scheduling [16], tool matching, and SPC for cycle time reduction and yield enhancement. Phase 3 (Mature) enjoys the economies of scale with excellent equipment reliability and mass production with low-cost and high-quality, particularly using big data volume that contributes to the virtual metrology (VM) [11], preventive/predictive maintenance (PM/PdM) [103], automated optical inspection (AOI) [104]. Figure 6 also encourages moving downstream labor force (e.g., quality assurance division) toward upstream to address R&D product design and ramping-up problems since uncertainties in the upstream bring difficulties but also opportunities for value creation.

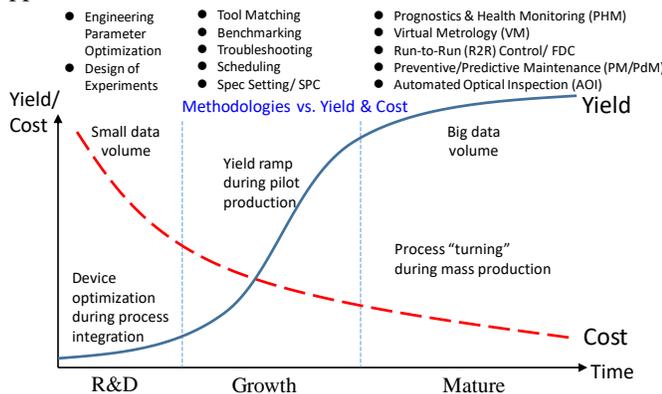

Fig. 6 Product development, yield, cost, and methodologies (revised from [92] [105])

For the future works of smart factory, a financial aspect can also provide insights. Figure 7 shows the decomposition of a key performance indicator (KPI) hierarchy related to return on equity (ROE). ROE measures business performance like the DuPont return on investment (ROI) formula [106] can be decomposed into three fundamental indices: profit margin which measures the manufacturer's ability to generate sales revenue and save cost, asset turnover which measures how effectively the manufacturer can generate revenue using asset investments, and financial leverage which measures the manufacturer's ability to use debt borrowed from stockholders to purchase assets. The first two indices can be associated with the manufacturing KPIs such as WIP, yield, cycle time, and bottleneck. Management's daily decisions on scheduling, product-mix, mean-time-to-repair (MTTR), mean-time-between-failure (MTBF), maintenance, etc., affect each KPI. These decisions coordinate and consume the manufacturing resources to improve the manufacturing system. The KPI hierarchy provides hints for building a big data platform to answer: how the daily decision in shop floor responses to our financial indices finally. The smart factory will benefit as data scientists develop new platforms for measuring the economic value of manufacturers' daily decisions.

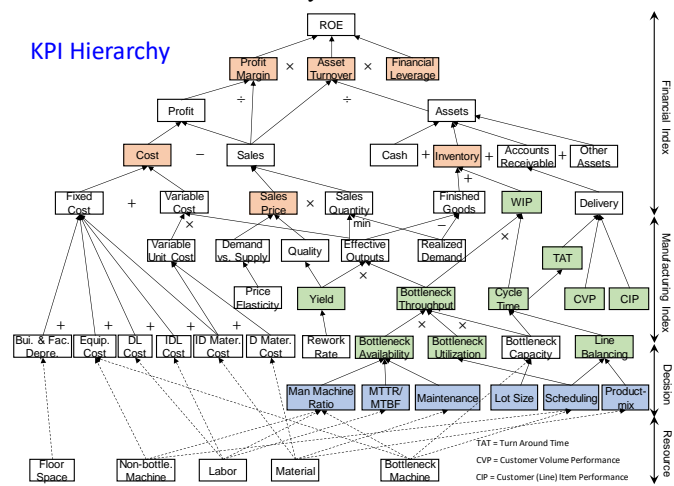

## Fig. 7 KPI hierarchy
## 3. Conclusion

This study describes the most common pitfalls and the potential solutions (protocols) encountered by data scientists in manufacturing systems. In fact, the answer to most of management issues in practice is "*it depends*" and thus it motivates this study to discuss the pitfalls and protocols. This study puts more focus on the practical aspect. For example, though ML/DS techniques provide powerful tools to solve the problems in real settings, ROI should be clarified in practice simultaneously to ensure a successful ML/DS project. Thus, it is critical to choose one substantial topic for improving the manufacturing system essentially and physically. In particular, the developed algorithm and analytics results should be embedded in the engineering automation system which is useful to realize our "ideas" substantially.

Finally, we present how the role of the manufacturing data science triggers the technology migration. Figure 8 illustrates the technology migration from big data to breeding data. The manufacturer evolves from MANUFACTURING(MFG).com, DATA.com, to INNOVATION.com and from problem-oriented, analytics-oriented, to idea-oriented. The three-phase iterative cycle triggers a technology migration to the next generation; in particular, data play an important role in this cycle to benefit the business growth and profitability. Furthermore, big data aggregated through the business evolution pushes the company moving forward and seeking a new business model to maintain the business core competence. Even similar manufacturers can use data in different way and generate different types of new technologies. Thus, beyond the big data, a business starts breeding data just like we breed kids for the future hope.

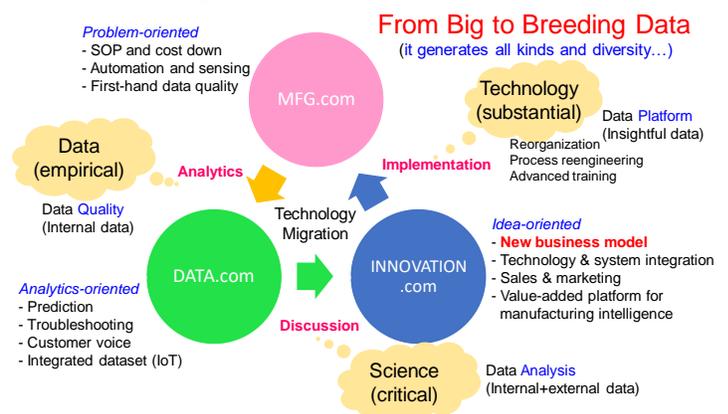

Fig. 8 Technology migration from big data to breeding data

## References

[1] J. Lee, E. Lapira, B. Bagheri, and H.-a. Kao, "Recent advances and trends in predictive manufacturing systems in big data environment," *Manufacturing Letters,* vol. 1, no. 1, pp. 38-41, 2013.

[2] J. Lee, B. Bagheri, and H.-A. Kao, "A Cyber-Physical Systems architecture for Industry 4.0-based manufacturing systems," *Manufacturing Letters,* vol. 3, pp. 18-23, 2015.

[3] G. P. Pisano, and S. C. Wheelwright, "The New Logic of High Tech R&D," *Harvard Business Review,* vol. 73, no. 5, pp. 93-105, 1995.

[4] T.-M. Choi, S. W. Wallace, and Y. Wang, "Big Data Analytics in Operations Management," *Production and Operations Management,* vol. 27, no. 10, pp. 1868-1883, 2018.

[5] C. Low, C.-M. Hsu, and K.-I. Huang, "Benefits of lot splitting in job-shop scheduling," *The International Journal of Advanced Manufacturing Technology,* vol. 24, no. 9-10, pp. 773-780, 2004.

[6] D. Golmohammadi, "A study of scheduling under the theory of constraints," *International Journal of Production Economics,* vol. 165, pp. 38-50, 2015.

[7] L. Shen, S. Dauzère-Pérès, and J. S. Neufeld, "Solving the flexible job shop scheduling problem with sequence-dependent setup times," *European Journal of Operational Research,* vol. 265, no. 2, pp. 503-516, 2018.

[8] C.-Y. Lee, and T.-L. Tsai, "Data science framework for variable selection, metrology prediction, and process control in TFT-LCD manufacturing," *Robotics and Computer Integrated Manufacturing,* vol. 55, pp. 76–87, 2019.

[9] W. J. Hopp, and M. L. Spearman, *Factory Physics*, 3 ed.: Waveland Press, 2011.

[10] Y.-T. Kao, S. Dauzère-Pérès, J. Blue, and S.-C. Chang, "Impact of integrating equipment health in production scheduling for semiconductor fabrication," *Computers & Industrial Engineering,* vol. 120, pp. 450-459, 2018.

[11] C. Fan-Tien, H. Hsien-Cheng, and K. Chi-An, "Developing an Automatic Virtual Metrology System," *IEEE Transactions on Automation Science and Engineering,* vol. 9, no. 1, pp. 181-188, 2012.

[12] W.-J. Lee, and S.-C. Ong, "Learning from small data sets to improve assembly semiconductor manufacturing processes."

[13] J. Lee, F. Wu, W. Zhao, M. Ghaffari, L. Liao, and D. Siegel, "Prognostics and health management design for rotary machinery systems—Reviews, methodology and applications," *Mechanical Systems and Signal Processing,* vol. 42, no. 1-2, pp. 314-334, 2014.

[14] S. Nahmias, and T. L. Olsen, *Production and Operations Analysis*, 7 ed.: Waveland Press, 2015.

[15] C.-Y. Lee, and C.-L. Liang, "Manufacturer's printing forecast, reprinting decision, and contract design in the educational publishing industry," *Computers & Industrial Engineering,* vol. 125, pp. 678-687, 2018.

[16] M. L. Pinedo, *Scheduling: Theory, Algorithms, and Systems*, 5 ed.: Springer, 2016.

[17] J.-Q. Wang, J. Chen, Y. Zhang, and G. Q. Huang, "Schedule-based execution bottleneck identification in a job shop," *Computers & Industrial Engineering,* vol. 98, pp. 308-322, 2016.

[18] C.-F. Chien, and C.-H. Chen, "A novel timetabling algorithm for a furnace process for semiconductor fabrication with constrained waiting and frequency-based setups," *OR Spectrum,* vol. 29, no. 3, pp. 391-419, 2007.

[19] W. Hung-Kai, C. Chen-Fu, and M. Gen, "An Algorithm of Multi-Subpopulation Parameters With Hybrid Estimation of Distribution for Semiconductor Scheduling With Constrained Waiting Time," *IEEE Transactions on Semiconductor Manufacturing,* vol. 28, no. 3, pp. 353-366, 2015.

[20] C.-Y. Lee, and A. L. Johnson, "Operational Efficiency," *Handbook of Industrial and Systems Engineering*, A. B. Badiru, ed., pp. 17–44: CRC Press, 2013.

[21] J. Liker, *The Toyota Way: 14 Management Principles from the World's Greatest Manufacturer*, 1 ed.: McGraw-Hill Education, 2004.

[22] S. Chopra, G. Reinhardt, and M. Dada, "The Effect of Lead Time Uncertainty on Safety Stocks," *Decision Sciences,* vol. 35, no. 1, pp. 1-24, 2004.

[23] R. Kohavi, "A study of cross-validation and bootstrap for accuracy estimation and model selection," *In: Proceedings of the International Joint Conference on Artificial Intelligence (IJCAI 1995),* pp. 1137-1143, 1995.

[24] L. V. Fausett, *Fundamentals of Neural Networks: Architectures Algorithms and Applications*, 1 ed.: Prentice Hall, 1994.

[25] Y. LeCun, Y. Bengio, and G. Hinton, "Deep learning," *Nature,* vol. 521, pp. 436–444, 2015.

[26] C. M. Bishop, *Pattern Recognition and Machine Learning*, 1 ed.: Springer, 2006.

[27] A. Liaw, and M. Wiener, "Classification and regression by randomforest," *R News,* vol. 2/3, pp. 18-22, 2002.

[28] J. H. Friedman, "Greedy function approximation: a gradient boosting machine," *Annals of Statistics,* vol. 29, no. 5, pp. 1189-1232, 2001.

[29] T. Hastie, R. Tibshirani, and J. Friedman, *The Elements of Statistical Learning: Data Mining, Inference, and Prediction*, 2 ed.: Springer, 2008.

[30] I. Guyon, and A. Elisseeff, "An introduction to variable and feature selection," *Journal of Machine Learning Research,* vol. 3, pp. 1157-1182, 2003.

[31] C.-Y. Lee, and B.-S. Chen, "Mutually-exclusive-and-collectively-exhaustive feature selection scheme," *Applied Soft Computing,* vol. 68, pp. 961-971, 2018.

[32] M. Efroymson, "Multiple regression analysis," *Mathematical Methods for Digital Computers,* vol. 1, pp. 191-203, 1960.

[33] L. Xu, and W.-J. Zhang, "Comparison of different methods for variable selection," *Analytica Chimica Acta,* vol. 446, no. 1-2, pp. 475-481, 2001.

[34] R. Tibshirani, "Regression shrinkage and selection via the lasso," *Journal of the Royal Statistical Society, Series B,* vol. 58, no. 1, pp. 267-288, 1996.

[35] L. Breiman, J. Friedman, R. A. Olshen, and C. J. Stone, *Classification and Regression Trees,* pp. CRC Press, 1984.

[36] R. Schapire, Y. Freund, P. Bartlett, and W. Lee, "Boosting the margin: A new explanation for the effectiveness of voting methods," *The Annals of Statistics,* vol. 26, no. 5, pp. 1651-1686, 1998.

[37] G. Rätsch, T. Onoda, and K. R. Müller, "Soft margins for adaboost," *Machine Learning,* vol. 42, pp. 287–320, 2001.

[38] K. Pearson, "On lines and planes of closest fit to systems of points in space," *The London, Edinburgh, and Dublin Philosophical Magazine and Journal of Science, Series 6* vol. 2, no. 11, pp. 559–572, 1901.

[39] C. Jutten, and J. Hérault, "Blind separation of sources, part I: an adaptive algorithm based on neuromimetic architecture," *Signal Processing,* vol. 24, pp. 1–10, 1991.

[40] P. Comon, "Independent component analysis, a new concept?," *Signal Processing,* vol. 36, pp. 287–314, 1994.

[41] J. H. Ward Jr., "Hierarchical grouping to optimize an objective function," *Journal of the American Statistical Association,* vol. 58, pp. 236–244, 1963.

[42] S. P. Lloyd, "Least square quantization in PCM," *Technical note, Bell laboratories, 1957. IEEE Transactions on Information Theory, 1982,* vol. 28, no. 2, pp. 129–137, 1957.

[43] C.-F. Chien, and S.-C. Chuang, "A framework for root cause detection of sub-batch processing system for semiconductor manufacturing big data analytics," *IEEE Transactions on Semiconductor Manufacturing,* vol. 27, no. 4, pp. 475-488, 2014.

[44] V. E. Velculescu, L. Zhang, B. Vogelstein, and K. W. Kinzler, "Serial analysis of gene expression," *Science,* vol. 270, no. 5235, pp. 484-487, 1995.

[45] Y. Saeys, I. Inza, and P. Larrañaga, "A review of feature selection techniques in bioinformatics," *Bioinformatics,* vol. 23, no. 9, pp. 2507–2517, 2007.

[46] D. C. Montgomery, *Design and Analysis of Experiments*, 8 ed.: Wiley, 2012.

[47] C. J. C. Burges, "A Tutorial on Support Vector Machines for Pattern Recognition," *Data Mining and Knowledge Discovery,* vol. 2, no. 2, pp. 121–167, 1998.

[48] D. F. Specht, "A general regression neural network," *IEEE Transactions on Neural Networks,* vol. 2, no. 6, pp. 568-576, 1991.

[49] T. Hastie, R. Tibshirani, and J. Friedman, *The Elements of Statistical Learning: Data Mining, Inference, and Prediction*, 2 ed.: Springer, 2009.

[50] W. Samek, T. Wiegand, and K.-R. Müller, "Explainable artificial intelligence: understanding, visualizing and interpreting deep learning models," 2017. https://arxiv.org/abs/1708.08296

[51] S. M. Lundberg, and S.-I. Lee, "A unified approach to interpreting model predictions," in 31st Conference on Neural Information Processing Systems (NIPS 2017), Long Beach, CA, USA., 2017.

[52] G. P. Zhang, "Time series forecasting using a hybrid ARIMA and neural network model," *Neurocomputing,* vol. 50, pp. 159-175, 2003.

[53] J.-L. Zhang, Y.-J. Zhang, and L. Zhang, "A novel hybrid method for crude oil price forecasting," *Energy Economics,* vol. 49, pp. 649-659, 2015.

[54] N. E. Huang, Z. Shen, S. R. Long, M. C. Wu, H. H. Shih, Q. Zheng, N.-C. Yen, C. C. Tung, and H. H. Liu, "The empirical mode decomposition and the Hilbert spectrum for nonlinear and non-stationary time series analysis," *Royal Society of London Proceedings Series A,* vol. 454, no. 1971, pp. 903-998, 1998.

[55] S. Beveridge, and C. R. Nelson, "A new approach to decomposition of economic time series into permanent and transitory components with particular attention to measurement of the 'business cycle'," *Journal of Monetary Economics,* vol. 7, no. 2, pp. 151-174, 1981.

[56] G. E. P. Box, and D. A. Pierce, "Distribution of residual autocorrelations in autoregressive-integrated moving average time series models " *Journal of the American Statistical Association,* vol. 65, no. 332, pp. 1509-1526, 1970.

[57] R. A. Johnson, and D. W. Wichern, *Applied Multivariate Statistical Analysis*, 6 ed.: Pearson, 2007.

[58] E. Keogh, S. Chu, D. Hart, and M. Pazzani, "Segmenting time series: a survey and novel approach," *In: Last, M., Kandel, A., Bunke, H. (eds.) Data Mining in Time Series Databases,* vol. 57, pp. 1-22, 2004.

[59] C.-Y. Lee, T.-S. Huang, M.-K. Liu, and C.-Y. Lan, "Data Science for Vibration Heteroscedasticity and Predictive Maintenance of Rotary Bearings," *Energies,* vol. 12, no. 5, 2019.

[60] S. Smith, *Digital Signal Processing: A Practical Guide for Engineers and Scientists*, 1 ed.: Newnes, 2003.

[61] D. B. Suits, "Use of dummy variables in regression equations," *Journal of the American Statistical Association,* vol. 52, no. 280, pp. 548–551, 1957.

[62] J. Han, M. Kamber, and J. Pei, *Data Mining: Concepts and Techniques*, pp. 3rd edition, Morgan Kaufmann, 2012.

[63] T. D. Pigott, "A review of methods for missing data," *Educational Research and Evaluation,* vol. 7, no. 4, pp. 353-383, 2001.

[64] A. R. Donders, G. J. van der Heijden, T. Stijnen, and K. G. Moons, "Review: a gentle introduction to imputation of missing values," *Journal of Clinical Epidemiology,* vol. 59, no. 10, pp. 1087-1091, Oct, 2006.

[65] R. M. O'Brien, "A caution regarding rules of thumb for variance inflation factors," *Quality & Quantity,* vol. 41, no. 5, pp. 673–690, 2007.

[66] J. F. Hair Jr, W. C. Black, B. J. Babin, and R. E. Anderson, *Multivariate Data Analysis*, 7 ed.: Prentice Hall, Upper Saddle River., 2009.

[67] G. James, D. Witten, T. Hastie, and R. Tibshirani, *An Introduction to Statistical Learning: with Applications in R*, 1 ed.: Springer, 2013.

[68] A. E. Hoerl, and R. W. Kennard, "Ridge regression: based eimation for nnorthogonal poblems," *Technometrics,* vol. 12, no. 1, pp. 55-67, 1970.

[69] J. N. Morgan, and J. A. Sonquist, "Problems in the analysis of survey data, and a proposal," *Journal of the American Statistical Association,* vol. 58, no. 302, pp. 415-434, 1963.

[70] G. V. Kass, "An exploratory technique for investigating large quantities of categorical data," *Journal of the Royal Statistical Society. Series C (Applied Statistics),* vol. 29, no. 2, pp. 119-127, 1980.

[71] J. R. Quinlan, "Induction of decision trees," *Machine learning,* vol. 1, pp. 81-106, 1986.

[72] J. R. Quinlan, *C4.5: Programs for Machine Learning* pp. Morgan Kaufmann Publishers Inc. San Francisco, CA, USA, 1993.

[73] R. T. Clemen, and T. Reilly, *Making Hard Decisions with Decision Tools*, 3 ed.: Cengage Learning, 2013.

[74] F. Hillier, and G. J. Lieberman, *Introduction to Operations Research*, 10 ed.: McGraw-Hill, 2015.

[75] L. J. Savage, "The theory of statistical decision," *Journal of the American Statistical Association,* vol. 46, pp. 55–67, 1951.

[76] J. R. Birge, and F. Louveaux, *Introduction to Stochastic Programming*, 2 ed.: Springer Verlag, New York, 2011.

[77] C.-Y. Lee, and M.-C. Chiang, "Aggregate demand forecast with small data and robust capacity decision in TFT-LCD manufacturing," *Computers & Industrial Engineering,* vol. 99, pp. 415-422, 2016.

[78] C.-W. Chou, C.-F. Chien, and M. Gen, "A Multiobjective Hybrid Genetic Algorithm for TFT-LCD Module Assembly Scheduling," *IEEE Transactions on Automation Science and Engineering,* vol. 11, no. 3, pp. 692-705, 2014.

[79] S.-H. Lin, C.-Y. Lee, T. Yang, and J.-C. Lu, "Two-phase simulation optimization for vendor selection and order allocation in a solar cell manufacturer.."

[80] H. R. Kerzner, *Project Management: A Systems Approach to Planning, Scheduling, and Controlling*, 12 ed.: John Wiley & Sons, Inc., Hoboken, New Jersey, 2017.

[81] V. Pillac, M. Gendreau, C. Guéret, and A. L. Medaglia, "A review of dynamic vehicle routing problems," *European Journal of Operational Research,* vol. 225, no. 1, pp. 1-11, 2013.

[82] H. M. Markowitz, "Portfolio Selection," *The Journal of Finance,* vol. 7, no. 1, pp. 77–91, 1952.

[83] C.-Y. Lee, and C.-F. Chien, "Stochastic programming for vendor portfolio selection and order allocation under delivery uncertainty," *OR Spectrum,* vol. 36, no. 3, pp. 761-797, 2014.

[84] R. E. Walpole, R. H. Myers, S. L. Myers, and K. E. Ye, *Probability and Statistics for Engineers and Scientists*, 9 ed.: Pearson, 2012.

[85] J. Reichertz, "Induction, Deduction, Abduction," *The SAGE Handbook of Qualitative Data Analysis*, U. Flick, ed.: SAGE Publications Ltd, 2014.

[86] J. A. Hoffer, R. Venkataraman, and H. Topi, *Modern Database Management*, 12 ed.: Pearson, 2015.

[87] A. B. M. Moniruzzaman, and S. A. Hossain, "Nosql database: New era of databases for big data analytics-classification, characteristics and comparison," *International Journal of Database Theory and Application,* vol. 6, no. 4, pp. 1-14, 2013.

[88] A. Freivalds, and B. Niebel, *Niebel's Methods, Standards, & Work Design*, 13 ed.: McGraw-Hill Education, 2013.

[89] R. L. Francis, L. F. McGinnis Jr., and J. A. White, *Facility Layout and Location: An Analytical Approach*, 2 ed.: Prentice-Hall, 1992.

[90] M. Hammer, and J. A. Champy, *Reengineering the Corporation: A Manifesto for Business Revolution*, 1 ed.: HarperBusiness, 1993.

[91] C. L. Philip Chen, and C.-Y. Zhang, "Data-intensive applications, challenges, techniques and technologies: A survey on Big Data," *Information Sciences,* vol. 275, pp. 314-347, 2014.

[92] Y.-C. Lin, M.-H. Hung, H.-C. Huang, C.-C. Chen, H.-C. Yang, Y.-S. Hsieh, and F.-T. Cheng, "Development of Advanced Manufacturing Cloud of Things (AMCoT)—A Smart Manufacturing Platform," *IEEE Robotics and Automation Letters,* vol. 2, no. 3, pp. 1809-1816, 2017.

[93] E. Sisinni, A. Saifullah, S. Han, U. Jennehag, and M. Gidlund, "Industrial Internet of Things: Challenges, Opportunities, and Directions," *IEEE Transactions on Industrial Informatics,* vol. 14, no. 11, pp. 4724-4734, 2018.

[94] X. Xu, "From cloud computing to cloud manufacturing," *Robotics and Computer-Integrated Manufacturing,* vol. 28, no. 1, pp. 75-86, 2012.

[95] J. Gama, I. Žliobaité, A. Bifet, M. Pechenizkiy, and A. Bouchachia, "A survey on concept drift adaptation," *ACM Computing Surveys,* vol. 46, no. 4, pp. 1-37, 2014.

[96] A. Rai, R. Patnayakuni, and N. Seth, "Firm Performance Impacts of Digitally Enabled Supply Chain Integration Capabilities," *MIS Quarterly,* vol. 30, no. 2, pp. 225-246, 2006.

[97] B. B. Flynn, B. Huo, and X. Zhao, "The impact of supply chain integration on performance: A contingency and configuration approach," *Journal of Operations Management,* vol. 28, no. 1, pp. 58-71, 2010.

[98] M. A. Waller, and S. E. Fawcett, "Data Science, Predictive Analytics, and Big Data: A Revolution That Will Transform Supply Chain Design and Management," *Journal of Business Logistics,* vol. 34, no. 2, pp. 77-84, 2013.

[99] S. Tiwari, H. M. Wee, and Y. Daryanto, "Big data analytics in supply chain management between 2010 and 2016: Insights to industries," *Computers & Industrial Engineering,* vol. 115, pp. 319-330, 2018.

[100] M. Brettel, N. Friederichsen, M. Keller, and M. Rosenberg, "How Virtualization, Decentralization and Network Building Change the Manufacturing Landscape: An Industry 4.0 Perspective," *International Journal of Information and Communication Engineering,* vol. 8, no. 1, pp. 37-44, 2014.

[101] T. Stock, and G. Seliger, "Opportunities of Sustainable Manufacturing in Industry 4.0," *Procedia CIRP,* vol. 40, pp. 536-541, 2016.

[102] E. Rauch, C. Linder, and P. Dallasega, "Anthropocentric perspective of production before and within Industry 4.0," *Computers & Industrial Engineering,* 2019.

[103] G. A. Susto, A. Schirru, S. Pampuri, S. McLoone, and A. Beghi, "Machine Learning for Predictive Maintenance: A Multiple Classifier Approach," *IEEE Transactions on Industrial Informatics,* vol. 11, no. 3, pp. 812-820, 2015.

[104] S.-H. Huang, and Y.-C. Pan, "Automated visual inspection in the semiconductor industry: A survey," *Computers in Industry,* vol. 66, pp. 1-10, 2015.

[105] J. Bordelon, and P. Maniar. "The sub-100-nm imperative: parametric yield ramp. EDN Network," May 6, 2019; https://www.edn.com/Home/PrintView?contentItemId=4018429.

[106] D. Brown, "Centralized control with decentralized responsibilities," *American Management Association Annual Convention,* vol. Series 57, pp. (reprinted in Johnson, H.T. (Ed.), Systems and Profits: Early Management Accounting at DuPont and General Motors (Arno Press, 1980)), 1927.